\documentclass[aps,prl,twocolumn,superscriptaddress]{revtex4}
\usepackage{graphicx}
\usepackage{xspace}

\begin{document}

\title{Fully anisotropic superconducting transition in ion-irradiated
YBa$_2$Cu$_3$O$_{7-\delta}$ with a tilted magnetic field}


\author{B. Espinosa-Arronte}
\email{beatriz@kth.se}
\author{M. Andersson}
\affiliation{Dept. of Microelectronics and Applied Physics, KTH 
School of ICT, Royal Institute of Technology, SE-164 40 Kista, Sweden}

\author{C. J. van der Beek}
\affiliation{Laboratoire des Solides Irradi\'{e}s, CNRS-UMR 7642 \& 
CEA/DSM/DRECAM, Ecole Polytechnique, F-91128 Palaiseau, France}

\author{M. Nikolaou}
\author{J. Lidmar}
\author{M. Wallin}
\affiliation{Dept. of Theoretical Physics, KTH School of Engineering 
Sciences, Royal Institute of Technology, SE-106 91 Stockholm, Sweden}

\date{\today}
\vspace{5cm}
\begin{abstract}

We consider the superconducting vortex solid-to-liquid transition
in heavy ion-irradiated untwinned YBa$_2$Cu$_3$O$_{7-\delta}$ single crystals 
in the case where the magnetic field direction does not coincide with that 
of the irradiation-induced linear columnar defects. For a certain 
range of angles, the resistivities measured in three orthogonal spatial directions 
vanish at the transition as three clearly different powers of reduced temperature. At previously known second-order phase transitions, scaling of physical quantities has either been isotropic or anisotropic in \emph{one} direction. Thus, our findings yield evidence for a 
new type of critical behavior with \emph{fully anisotropic} critical 
exponents. 

\end{abstract}


\maketitle



Vortex matter in high temperature superconductors offers many opportunities to study the effects of disorder, fluctuations, and frustration, which lead to a variety of new phases and phase transitions~\cite{Blatter94}. Among these, the two principal thermodynamic phases are a dissipative vortex liquid at high temperatures, and a truly superconducting vortex solid at low 
temperatures. In clean superconductors, the vortex solid-to-liquid transition is a first-order melting transition.  The presence of strong disorder is believed to turn it into a continuous isotropic ``vortex glass'' transition~\cite{Fisher91}. Then, many quantities develop power-law singularities upon approaching the transition, which define its critical exponents. The critical properties are insensitive to microscopic details such as small-scale anisotropies,
and depend instead on such general features as symmetry and dimensionality, which determine the transition's universality class. All phase transitions discovered so far are of two main types, those with isotropic scaling properties, and those with anisotropic scaling in one direction.

For point disorder the glass transition will be \emph{isotropic}, {\em i.e.}, the critical exponents are independent of direction, even though the material itself can be anisotropic, as is the case in, {\em e.g.},  YBa$_{2}$Cu$_{3}$O$_{7-\delta}$ (YBCO). A markedly different situation occurs in superconductors containing linear 
columnar defects following irradiation by swift heavy ions~\cite{Nelson9293,Nelson96,Lidmar99,Jiang94,Grigera98,Reed95,Olsson02}.
Such defects are very effective at pinning the vortex lines
when a magnetic field is applied parallel to them; the vortex solid at 
low temperature is then a ``Bose glass''~\cite{Nelson9293}.  The columnar defects and
the applied field both break the symmetry, making the Bose glass
transition \emph{anisotropic}, with different exponents 
parallel and perpendicular to the columns~\cite{Nelson9293}.
An interesting situation occurs when the magnetic field is tilted away
from the defects.  For small tilt the vortices will stay on
the columns, leading to a transverse Meissner effect.  As the tilt
angle or the temperature is increased the vortices will eventually
depin and enter the vortex liquid phase.  The transverse component of
the magnetic field then makes all three directions nonequivalent and
opens for the possibility of fully anisotropic scaling, i.e., with different critical exponents in all directions~\cite{Vestergren05}.  Furthermore, the problem of vortex
depinning from columnar defects is formally equivalent to the
Bose-glass transition of bosons in a random
potential~\cite{Nelson9293}.  In this analogy, the tilted magnetic
field corresponds to an imaginary vector potential, resulting in a
non-Hermitian localization problem~\cite{Hatano96}.

Here, we investigate the vortex solid-to-liquid transition 
by electrical transport measurements performed in three orthogonal spatial 
directions on untwinned heavy-ion irradiated YBCO with the magnetic field 
tilted away from the columns. It appears that, for angles not too close
to the principal crystal axes, the resistivity vanishes as a power law of 
reduced temperature with different exponents in the three directions. 
Thus, we provide experimental evidence for a phase transition with anisotropic critical exponents in all directions, thereby extending the possible type of universality classes.

YBCO single crystals were grown by a self-flux method in yttria 
stabilized zirconia crucibles~\cite{Eltsev94}. As-grown untwinned 
crystals with typical sizes of $0.5 \times 0.2 \times 0.02$~mm$^{3}$ 
were annealed at 400~$^{\text{o}}$C in flowing O$_2$ for a week. Such crystals show clear vortex melting signatures in magnetic fields. Irradiation with 1 GeV Pb$^{56+}$ ions was performed at the Grand Acc\'{e}l\'{e}rateur National d\'{}Ions Lourds (GANIL), Caen, France, 
with the incident beam almost parallel to the crystallographic $c$-axis. The resulting damage consists of randomly distributed amorphous tracks, with a diameter of about 7~nm, extending through the total thickness of the sample \cite{Yan98}. Every ion impact creates a track. For most samples, the track density was $n_{d} = 1 \times 10^{11}$ ions cm$^{-2}$, corresponding to a matching field $B_{\phi} \equiv \phi_{0} / n_{d} = 2.0$ T ( $\phi_{0} = h/2e$  
is the flux quantum). Electrical contacts were prepared by silver paint, giving contact  
resistances below 1.5 $\Omega$. For resistivity measurements in the $x$ and $y$ directions,
perpendicular to the columns and parallel to the CuO$_{2}$ planes, the current pads were applied on opposite sides of the crystal. The resistivity parallel to the columns, $\rho_{z}$, was measured on another sample, with current contacts covering most of the ab-plane surfaces so as to assure the most homogeneous current distribution~\cite{Lundqvist01}. The linear resistivity was measured using the standard four-probe technique, with a current of 0.1 mA and 1 mA for the in-plane and $c$-axis measurements respectively. Using a dc picovoltmeter as preamplifier, the voltage resolution was below 1 nV. The angular resolution of the field orientation is 0.01$^{\circ}$ along the direction of rotation of our single-axis rotating sample holder and 1-2$^{\circ}$ perpendicular to it.



\begin{figure}
\includegraphics{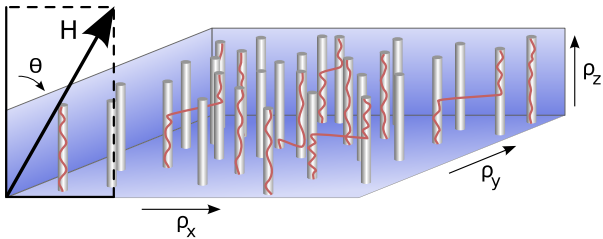}
\caption{(Color online). Experimental geometry. The columnar defects 
are parallel to the $z$~direction, which coincides with the 
crystallographic $c$~axis. The magnetic field is always applied in the 
$xz$--plane, at an angle~$\theta$ with respect to $z$. Vortex lines may accommodate to the columns by forming kinks.} 
\label{fig:geometry}
\end{figure}

\begin{figure}
\includegraphics{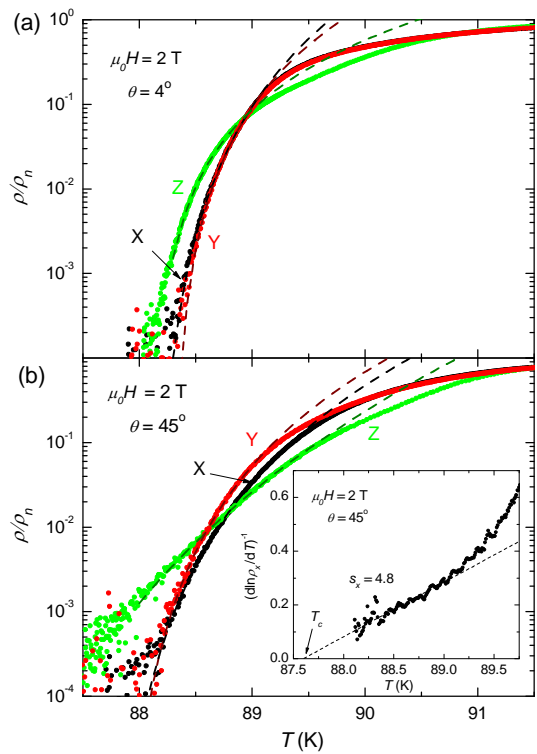}
\caption{(Color online). Normalized resistivity in three 
independent directions of an untwinned heavy-ion irradiated YBCO 
single crystal with a matching field $B_{\phi}=2$~T, and at tilt angles (a)~$\theta=4^{\circ}$ and (b)~$\theta=45^{\circ}$ from the columns. Dashed lines are fits 
to~Eq.~(\ref{eq:glass}). Inset: Extraction of $T_c$ and the critical 
exponent $s_i$ using $(\partial \ln \rho_{i} / \partial T)^{-1} = 
(T-T_{c})/s_{i}$.} \label{fig:resistivity}
\end{figure}

The geometry of our experiment is shown in Fig.~\ref{fig:geometry}. The magnetic field, $H$, is always applied in the $xz$ plane at an angle $\theta$ with respect to the direction of the columnar defects; $\rho_{x,y}$ were measured on the same sample mounted in different orientations with respect to the field direction. Fig.~\ref{fig:resistivity}~shows raw $\rho (T)$  data measured in the three spatial directions. The zero-field transition temperature for the crystal used for most of the  $\rho_{x}$ and  $\rho_{y}$~measurements was $T_{c0}=91.7$~K ($\Delta T_{c}\approx 0.5$~K); the crystal employed to measure $\rho_z$ had $T_{c0}=91.2$~K ($\Delta T_{c}\approx 1$~K). The difference in $T_{c0}$ accounts for the different vortex solid-to-liquid transition temperature $T_{c}$ found in the $\rho_z$ measurements. Our results are independent of the choice of $x$ as either the crystallographic $a$ or $b$~axis.

Close to the vortex solid-to-liquid phase transition, the resistivity 
drops to zero as a power law of $\left| T/T_{c} - 1 \right|$,
\begin{equation}	
\rho_{i} = \rho_{0i} \left| T/T_{c} - 1 \right|^{s_{i}};\; i=x, y, z.	
\label{eq:glass}
\end{equation}
From the measured $\rho_{i} (T)$ curves, $s_i$ and $T_{c}$ were determined in the standard way. A straight line was fit to the inverse logarithmic derivative of the experimental data, $(\partial \ln \rho_{i} / \partial T)^{-1} = (T-T_{c})/s_{i}$, as shown in~Fig.~\ref{fig:resistivity}. Below 5\% of the normal state resistivity, power-law scaling corresponding to the critical regime is observed. All measurements are made in the linear response regime; therefore, the scaling is insensitive to circumstances such as possible non-locality of vortex flow. We have also directly fitted the data to a power law (see Fig.~\ref{fig:resistivity}), which yields consistent results.



\begin{figure}
\includegraphics[scale=1]{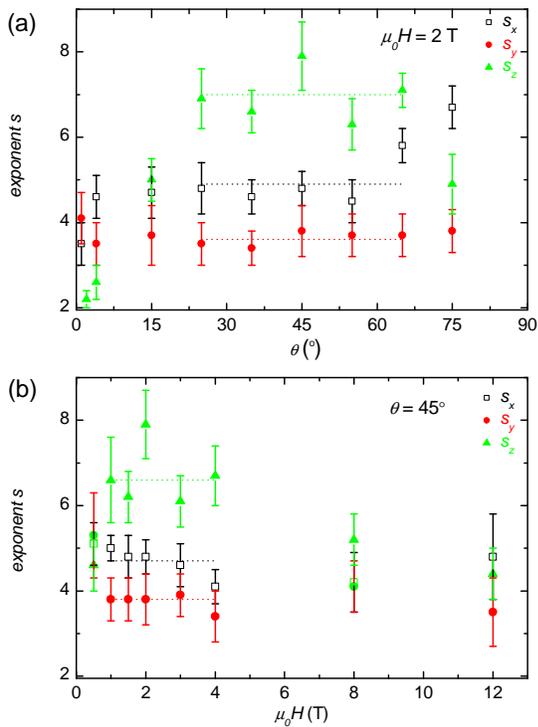}
\caption{(Color online). Exponents $s_{x}$, $s_y$ and $s_z$ of the 
power-law Eq.\protect(\ref{eq:glass}) by which the resistivity vanishes at the vortex solid-to-liquid 
transition in the $x$, $y$ and $z$ directions respectively. 
(a)~Exponents at 2~T as function of angle $\theta$.  
For 15$^{\circ} \lesssim \theta \lesssim $65${^\circ}$, the 
scaling behavior is anisotropic in three directions. (b)~Exponents at 
$\theta = 45^{\circ}$, as function of magnetic field. There is anisotropic 
scaling at fields close to the matching field, $\mu_{0}H \approx B_{\phi}=2$~T. Dotted 
lines show the average value of $s_i$ in the regimes of anisotropic scaling. 
The error bars are estimated from the quality of the 
fits in Fig.\ref{fig:resistivity}.
\label{fig:exponents}}
\end{figure}

The experimentally obtained exponents $s_{x}$, $s_{y}$ and $s_{z}$ 
for $\mu_0 H = B_{\phi}$ are shown  as function of $\theta$ in 
Fig.~\ref{fig:exponents}(a). There are three distinct angular 
regimes, with different relations between the three 
exponents. For $\theta \lesssim 15^{\circ}$, $s_x \approx s_y$, 
while $s_{z}$ is smaller. For $15^{\circ}\lesssim \theta \lesssim 65^{\circ}$  
all $s_i$ are different, with $s_{y} < s_{x} < s_{z}$. In this interval       
the best fits to the power law (\ref{eq:glass}) were found. The exponents 
are approximately constant as a function of angle, suggesting a certain  
amount of universality. The average values $s_{x}=4.9\pm 0.2$, 
$s_{y}= 3.6\pm 0.2$, and $s_{z}=6.9\pm 0.3$ are indicated by the dashed 
lines in Fig.~\ref{fig:exponents}(a). Finally, at angles above 
65$^{\circ}$, $s_z$ decreases while $s_x$ increases. The error bars 
were estimated by varying the resistance interval of the fits while 
maintaining good agreement with the data.  

\begin{figure}
\includegraphics{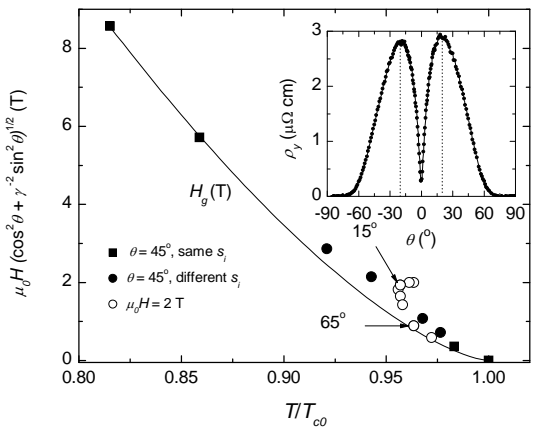}
\caption{Phase diagram for untwinned heavy-ion irradiated YBCO single 
crystals. An effective magnetic field is used to account for the 
anisotropy of YBCO. The symbols show the vortex solid-to-liquid 
transition temperatures obtained from measurements of $\rho_{y}$ at 
constant $\theta=45^{\circ}$ and several fields 
(\protect\rule{1.5mm}{1.5mm},$\bullet$); and at constant $\mu_{0}H=2$~T and 
different $\theta$ ($\circ$). At fields far from $B_{\phi} = 2$ T,
 isotropic glass exponents are found (\protect\rule{1.5mm}{1.5mm}) and 
the data can be fitted to Eq.~(\protect\ref{eq:glassline}) (solid line). For fields 
comparable to $B_{\phi}$ the transition 
lies above this line. At $\theta \approx 15^{\circ}$ the transition bends 
towards the isotropic glass line; nevertheless, it remains above it up to $\theta \sim  
65^{\circ}$. Inset: $\rho_y (\theta)$ at 1~T and 90.2~K. At 
$\theta=0$ the field is aligned with the columns. The 
accommodation angle $\theta_{a}$ is usually defined by the maximum of 
$\rho(\theta)$ (dashed lines). } \label{fig:scalediagram}
\end{figure}

The interaction between the columns and the vortices changes with tilt
angle, which can explain the existence of several angular regimes with
different behavior of the exponents. At angles below 
$\theta_{L}\approx 1-2^{\circ}$~\cite{Smith00}, the vortices
are locked within the columns in the Bose-glass phase and a transverse
Meissner effect is expected~\cite{Nelson9293,Smith00}. The proximity
to this phase will influence the transition and may explain the
variation in the exponents for $\theta <
5^{\circ}$~\cite{Smith00}. For $\theta > \theta_{L}$, vortices adjust
to the columnar defects by forming a staircase
structure~\cite{Blatter94,Nelson9293}, depicted in
Fig.~\ref{fig:geometry}. This occurs up to a certain accommodation
angle $\theta_a$ determined by the balance between the pinning energy
gained by the segments trapped on a column and the energy loss due to
 vortex line tilt. For $\theta > \theta_{a}$, the columns are
thought to act as translationally symmetric ``point pins''. The vortex
lines intersect the columns only over a length corresponding to the
vortex core radius, and are on average parallel to the direction of
the applied field. Experimentally, $\theta_a$ is usually defined as
the maximum of the angular-dependent
resistivity~\cite{Budhani94,Pomar01}, as shown in the inset of
Fig.~\ref{fig:scalediagram}. 
In our experiments this gives $\theta_{a}
\approx 15-20^{\circ}$, which coincides with the observed change in
the behavior of the critical exponents.  However, it is \emph{above}
this so defined $\theta_{a}$ that three different resistivity
exponents are observed. This suggests that the columnar defects still
play a determining role in this regime, as they break the symmetry
sufficiently strongly for the anisotropic scaling to be observed. Finally, as the field orientation perpendicular to the columns is approached, another change in the sequence of the exponents is observed, probably due to the layered nature of YBCO~\cite{Kwok94intrinsic}.


Fig.~\ref{fig:exponents}(b) shows that the values of $s_i$ change as a function of $H$,
indicating a change in the nature of the phase transition. At high fields, 
the vortex density exceeds the defect density. Hence,  
vortex-vortex interactions become increasingly important compared to 
the vortex-defect interaction, and the anisotropy of the transition 
described above should progressively disappear.  Above 4~T, the exponents are 
roughly the same in all directions and more or less field 
independent. The difference between the exponents is most pronounced 
close to the matching field. The average values in this region are 
$s_{x}=4.7\pm 0.2$, $s_{y}=3.8\pm 0.2$ and $s_{z}=6.6 \pm 0.3$, 
consistent with the average values at constant $\mu_{0}H=2$~T found 
above. At lower fields, the exponents again approach a common value. We observe these trends systematically irrespective of sample or matching field.

For high fields the effect of the columns is less relevant. Most 
vortices are pinned by pointlike pinning centers, or by the 
interaction with the ``matrix'' of vortices trapped on a 
column~\cite{Menghini03}. The vortex solid resembles an isotropic 
glass, with identical power-law exponents in all directions. Then, 
the glass line is given by \cite{Klein98}, 
\begin{equation}
H_{g}(T)\propto (1-T/T_{c0})^n, \label{eq:glassline}
\end{equation}
where $T_{c0}$ is the zero-field transition temperature and 
$H_{g}=H(T=T_{c}(H))$. Fig.~\ref{fig:scalediagram} shows the 
experimentally obtained solid-to-liquid transition temperatures 
$T_c(H)$ obtained from $\rho_{y}$. To account for the intrinsic 
anisotropy of YBCO an effective field 
$H_{\text{eff}}=H \sqrt{\gamma^{-2}\sin^{2}\theta+\cos^{2}\theta} $ 
is plotted on the ordinate ( $\gamma \approx 7$ is the anisotropy 
parameter~\cite{Blatter92}). The Figure also shows a fit to Eq.~(\ref{eq:glassline})
of the $T_c$ data at fields where approximately isotropic exponents $s_i$ were found. The fit gives $n\approx 1.5$, in agreement with the value reported for the 
isotropic vortex glass~\cite{Klein98}. We observe that for fields close 
to the matching field, and angles $\theta \lesssim 65^{\circ}$, $T_{c}(H)$ 
lies above the glass line (\ref{eq:glassline}). A bend towards this line is observed at an 
angle consistent with the experimentally defined $\theta_{a}\approx 
15^{\circ}$ but, up to $\theta \sim 65^{\circ}$, the data still lie 
above the $H_{g}(T)$ line. In fact, $T_{c}(H)$ resembles the behavior 
expected for the Bose-glass line~\cite{Krusin-Elbaum94, Blatter94, 
Larkin95}, adding further evidence that the effect of the 
columns as correlated pinning centers is important even at angles 
{\em above} the experimentally defined $\theta_a$. 

We now discuss the value of the exponents in the field- ($ B\simeq B_{\phi}$) and angular regime ($15^{\circ} < \theta < 65^{\circ}$) in which fully anisotropic scaling is observed. We assume that the anisotropic scaling is due to the different correlation length exponents in the three directions. Then, $\xi_x \sim \xi^\chi$, $\xi_y \sim \xi$ and $\xi_z \sim \xi^\zeta$, where $\xi \sim |T-T_c|^{-\nu}$. We also assume dynamic scaling for critical slowing down, described by the correlation time $\tau \sim \xi^z$, where $z$ is the dynamic critical exponent. Following the procedure outlined in Ref.~\cite{Vestergren05}, the longitudinal resistivities are $\rho_{i} = \frac {E_i}{J_i} \sim (T-T_{c})^{s_i}$,
where $s_x = \nu (z-1+\chi-\zeta)$, $s_y = \nu (z+1-\chi-\zeta)$ and 
$s_z = \nu (z-1-\chi+\zeta)$. Thus, the observation of three 
different $s_i$'s suggests that the correlation length diverges with 
different exponents in all directions. An additional constraint is 
needed for extracting $\nu$, $\chi$ and $\zeta$ from the experiments. 
Given the similarity between the Bose glass and the transverse Meissner transitions~\cite{Vestergren05} on the one hand and the transition observed here on the other, one may argue that the relation $\zeta - \chi -1 = 0$, which follows if the compressibility
remains finite at the transition, should remain valid. If this is so, 
the experimentally observed $s_i$ yield $\nu = 1.0 \pm 0.4$, $\chi = 1.6
\pm 0.7$, $\zeta = 2.6 \pm 0.7$ and $z=6.9 \pm 2.2$. This can be compared with the results
on the transverse Meissner transition ($\nu \approx 0.70$, $\chi = 2$
and $\zeta = 3$), indicating that the transition observed here may be
of somewhat different nature.

In summary, for certain angles between the field and the columns 
and at vortex densities close to the defect density, the resistivity 
of heavy-ion irradiated untwinned YBCO vanishes as a power law of 
reduced temperature with different exponents in all three directions. 
This shows that the scaling properties of the phase transition 
are fully anisotropic in three dimensions. Thus, our findings provide 
experimental evidence for a new kind of universal scaling behavior at 
a second-order phase transition. 
It would be of interest to search for anisotropic
scaling behavior in other physical systems with a similar symmetry
breaking as the one considered here.

We thank M.~Konczykowski for his presence during the irradiation and \"O.~Rapp 
for valuable comments. This work has been supported by the Swedish 
Research Council (Vetenskapsr{\aa}det) and the Swedish Foundation for 
Strategic Research through the OXIDE program. The G\"oran Gustafsson 
Foundation is acknowledged for supporting the work of M.~N. and 
financing equipment.


\end{document}